\documentclass[12pt,a4paper]{article}
\usepackage{jheppub}

\author[a]{David Emmanuel-Costa,}
\author[a,b]{Edison T. Franco}
\author[c,a]{and Ricardo Gonz\'{a}lez Felipe}
\affiliation[a]{Departamento de F\'{\i}sica and Centro de F\'{\i}sica Te\'{o}rica de Part\'{\i}culas,\\
Instituto Superior T\'{e}cnico, Av. Rovisco Pais, 1049-001 Lisboa, Portugal}
\affiliation[b]{Instituto de F\'{\i}sica Gleb Wataghin, Universidade Estadual de Campinas,\\
 Rua S\'{e}rgio Buarque de Holanda 777, 13083-970, Campinas, SP, Brazil}
\affiliation[c]{Instituto Superior de Engenharia de Lisboa, Rua Conselheiro Em\'{\i}dio Navarro 1, 1959-007 Lisboa, Portugal}
\emailAdd{david.costa@ist.utl.pt}
\emailAdd{edison.franco@ist.utl.pt}
\emailAdd{gonzalez@cftp.ist.utl.pt}

\title{$\mathsf{SU(5)\times SU(5)}$ unification revisited}

\abstract{The idea of grand unification in a minimal supersymmetric $\mathsf{SU(5)\times SU(5)}$ framework is revisited. It is shown that the unification of gauge couplings into a unique coupling constant can be achieved at a high-energy scale compatible with proton decay constraints. This requires the addition of a minimal particle content at intermediate energy scales. In particular, the introduction of the $\mathsf{SU(2)_L}$ triplets belonging to the $(15,1)+(\overline{15},1)$ representations, as well as of the scalar triplet $\Sigma_3$ and octet $\Sigma_8$ in the $(24,1)$ representation, turns out to be crucial for unification. The masses of these intermediate particles can vary over a wide range, and even lie in the TeV region. In contrast, the exotic vector-like fermions must be heavy enough and have masses above $10^{10}$~GeV. We also show that, if the $\mathsf{SU(5)\times SU(5)}$ theory is embedded into a heterotic string scenario, it is not possible to achieve gauge coupling unification with gravity at the perturbative string scale.}


\begin{document}

\maketitle

\section{Introduction}
\label{sec:Introduction}

On the quest for the theory beyond the Standard Model (SM), supersymmetric grand unified theories (SUSY GUTs) have revealed many attractive features which can solve some of the aspects left unexplained in the SM. This idea is supported by the unification of the gauge couplings that occurs, through renormalization group evolution, at a scale around $10^{16}$~GeV in the minimal supersymmetric standard model (MSSM). In the latter case, the SUSY threshold is set in the TeV region.

Since the appearance of the simplest GUT models proposed in 1974 by Georgi and Glashow, and based in the gauge group $\mathsf{SU(5)}$~\cite{Georgi:1974sy}, the search for gauge groups compatible with a unification scheme has been actively pursued in the literature~\cite{Georgi:1975qb,Langacker:1980js,Raby:2008gh}. Yet the unification and breaking patterns are far from being established. The low-energy supersymmetric $\mathsf{SU(5)}$ version~\cite{Sakai:1981gr} has been quoted as an excellent unification theory, since in this model gauge couplings unify very precisely at one-loop level without the need of new particles. Moreover, at two-loop~\cite{Barger:1992ac} and three-loop~\cite{Martens:2010nm} levels, gauge unification can also be achieved if threshold effects are taken into account.

Besides being successful in unifying gauge couplings, GUTs should also address other theoretical challenges. The proton should live long enough~\cite{Buras:1977yy,Langacker:1980js,Sakai:1981pk,Hayato:1999az, Nath:2006ut}. This requirement usually leads to the well-known doublet-triplet splitting problem, i.e. the $\mathsf{SU(2)}_{L}$ doublet and the $\mathsf{SU(3)}_{C}$ colour triplet belonging to the same multiplet must have a strong mass hierarchy. In other words, the parameters in the Higgs potential responsible for the doublet and triplet masses must be highly fine tuned.

Going beyond the simplest $\mathsf{SU(5)}$ unification, it is also conceivable that the unification group has a semi-simple structure, as in the original left-right symmetric Pati-Salam model~\cite{Pati:1973rp,Pati:1974yy}. In this direction, the SUSY left-right $\mathsf{SU(5)}\times \mathsf{SU(5)}$ model~\cite{Davidson:1987mi,Dine:2002se} has many attractive features that are absent in minimal realizations of the $\mathsf{SU(5)}$ theory. Indeed, R-parity can be automatically conserved, proton decay is suppressed because heavy and light fermions do not mix, the doublet-triplet splitting problem is alleviated~\cite{Barr:1996kp,Dine:2002se}, a generalized seesaw mechanism for fermion masses can be easily incorporated, and nonvanishing neutrino masses are naturally explained. Furthermore, $\mathsf{SU(5)}\times \mathsf{SU(5)}$ theories can be easily embedded in superstring constructions~\cite{Mohapatra:1996fu,Mohapatra:1996iy} which aim at unifying gravity with electroweak and strong forces. In what concerns unification, it is worth noticing that the same discrete permutation symmetry that guarantees the left-right nature of $\mathsf{SU(5)}\times \mathsf{SU(5)}$ (i.e. the one-to-one correspondence among left and right matter field representations) also leads to the unification of gauge couplings into a single constant.

If one assumes that the $\mathsf{SU(5)}\times \mathsf{SU(5)}$ group breaks directly to the SM gauge group $\mathsf{SU(3)}_C \times \mathsf{SU(2)}_L \times \mathsf{U(1)}_Y$ at the unification scale $\Lambda$, then the three SM gauge couplings $g_a \, (a=s,w,y)$ meet together into a single value,
\begin{equation}\label{unifcond}
\alpha_U=k_3\,\alpha_s =k_2\,\alpha_w =k_1\,\alpha_y\,,
\end{equation}
where $\alpha_a = g_a^2/(4\pi)$. The coefficients $k_i$ are group factors,
$k_i = (\text{Tr}\, T^2_i) / (\text{Tr}\, T^2)$, $(i=1,2,3)$,
where $T$ and $T_i$ are generators of the GUT group properly normalized over the full group and its SM subgroup $G_i$, respectively. For $\mathsf{SU(5)}\times \mathsf{SU(5)}$ one obtains the non-canonical values $k_1=13/3, k_2 =1$ and $k_3=2$. The corresponding weak mixing angle at the unification scale is given by
\begin{equation}
\sin ^{2}\theta _{W} = \frac{\alpha_y}{\alpha_y+\alpha_w}=\frac{1}{1+k_{1}/k_{2}}=\frac{3}{16}\,.
\end{equation}
It is commonly believed that this value cannot be reconciled with measurements at the electroweak scale, since it is rather small and, in general, $\sin ^{2}\theta _{W}$ decreases from high to low energies~\cite{Cho:1993jb,Mohapatra:1996fu,Mohapatra:1996iy}. Yet, if some appropriate representations are taken into account in the renormalization group evolution of the gauge couplings, this may not be the case. In particular, we shall show that the inclusion of the $(\overline{15},1)+(1,15)$ and their conjugate $(15,1)+(1,\overline{15})$ representations is sufficient to drive $\sin ^{2}\theta _{W}$ to the correct value. This is due to the fact that the $\mathsf{SU(2)}_{L}$ triplets contained in the $15$ and $\overline{15}$ representation of $\mathsf{SU(5)}_{L}$ strongly adjust the $\alpha _{w}$ coupling constant. It is also remarkable that the above representations play a crucial role in implementing the seesaw mechanism for neutrino masses.

In this work we revive the idea of grand unification in the supersymmetric version of the left-right $\mathsf{SU(5)\times SU(5)}$ gauge group. Our aim is to demonstrate that, with the addition of a minimal particle content, it is possible not only to unify the SM gauge coupling constants into a single GUT value, but also to bring the theory into agreement with the electroweak observational data. The paper is organized as follows. In Sec.~\ref{sec:model} we introduce the particle content of the model and discuss possible breaking patterns to the SM gauge group. We also briefly address the question of fermion masses in the context of the generalized seesaw. The unification of gauge couplings at one-loop and two-loop levels is studied in~\ Sec.~\ref{sec:gut} and a general numerical analysis is presented in Sec.~\ref{sec:num}. Finally, our concluding remarks are given in Sec.~\ref{sec:Conclusions}.

\section{The model}
\label{sec:model}

The supersymmetric left-right $\mathsf{SU(5)\times SU(5)}$ gauge group contains two copies per generation of the usual SUSY $\mathsf{SU(5)}$ theory. In the left-handed picture, the $(\overline{5}+10,1)$ fermion representations, denoted by $\psi$ and $\chi$, are given by
\begin{equation}
\label{eq:ferm}
\psi =\left[
\begin{array}{c}
D_{1}^{c} \\
D_{2}^{c} \\
D_{3}^{c} \\
e \\
-\nu
\end{array}
\right] \sim (\overline{5},1),\quad \chi =\frac{1}{\sqrt{2}}\left[
\begin{array}{ccccc}
0 & U_{3}^{c} & -U_{2}^{c} & -u_{1} & -d_{1} \\
-U_{3}^{c} & 0 & U_{1}^{c} & -u_{2} & -d_{2} \\
U_{2}^{c} & -U_{1}^{c} & 0 & -u_{3} & -d_{3} \\
u_{1} & u_{2} & u_{3} & 0 & -E^{c} \\
d_{1} & d_{2} & d_{3} & E^{c} & 0
\end{array}
\right] \sim (10,1),
\end{equation}
while the $(1,5+\overline{10})$ fields, represented by $\psi^{c}$ and $\chi^{c}$, are
\begin{equation}
\label{eq:fermconj}
\psi ^{c}=\left[
\begin{array}{c}
D_{1} \\
D_{2} \\
D_{3} \\
e^{c} \\
-\nu ^{c}
\end{array}
\right] \sim (1,5),\quad \chi ^{c}=\frac{1}{\sqrt{2}}\left[
\begin{array}{ccccc}
0 & U_{3} & -U_{2} & -u_{1}^{c} & -d_{1}^{c} \\
-U_{3} & 0 & U_{1} & -u_{2}^{c} & -d_{2}^{c} \\
U_{2} & -U_{1} & 0 & -u_{3}^{c} & -d_{3}^{c} \\
u_{1}^{c} & u_{2}^{c} & u_{3}^{c} & 0 & -E \\
d_{1}^{c} & d_{2}^{c} & d_{3}^{c} & E & 0
\end{array}
\right] \sim (1,\overline{10}).
\end{equation}
The multiplets of Eqs.~(\ref{eq:ferm}) and (\ref{eq:fermconj}) have extra fermions beyond those present in the SM: the vector-like fermions ($U$,\,$U^c$,\,$D$,\,$D^c$,\,$E$,\,$E^c$) and the well-motivated right-handed neutrino, $\nu^{c}$. There is no vector-like analog of the neutrino.

To discuss the breaking scheme to the SM gauge group, one needs to specify the Higgs content. Among the different possibilities, here we consider the following pattern:
\begin{equation}
\label{eq:pattern}
\begin{array}{c}
\mathsf{SU(5)}_{L}\times \mathsf{SU(5)}_{R} \\
\downarrow \Lambda \\
\mathsf{SU(3)}_{L}\times \mathsf{SU(2)}_{L}\times \mathsf{U(1)}_{L}\times \mathsf{SU(3)}_{R} \times
\mathsf{SU(2)}_{R}\times \mathsf{U(1)}_{R} \\
\downarrow \Lambda_{LR} \\
\mathsf{SU(3)}_{C} \times \mathsf{SU(2)}_{L}\times \mathsf{SU(2)}_{R}\times \mathsf{U(1)}_{B-L}\\
\downarrow v_{R} \\
\mathsf{SU(3)}_{C}\times \mathsf{SU(2)}_{L}\times \mathsf{U(1)}_{Y}\\
\downarrow v_{L} \\
\mathsf{SU(3)}_{C}\times \mathsf{U(1)}_{em}\,.
\end{array}
\end{equation}
We identify $\mathsf{SU(3)}_{C}$ with the $\mathsf{SU(3)}_{L+R}$ diagonal subgroup and $\mathsf{U(1)}_{B-L}$ with $\mathsf{U(1)}_{L+R}\,$. The breaking energy scales $\Lambda, \Lambda_{LR}$ and $v_{R}$ are determined by the Higgs content of the model. In this implementation, we need the adjoint representations of both $\mathsf{SU(5)}$ subgroups. We introduce $\Phi _{L}\sim (24,1)$ and $\Phi _{R}\sim (1,24)$, which accomplish the first breaking of $\mathsf{SU(5)_{L}\times SU(5)_{R}}$ at the scale $\Lambda$ but preserve the discrete left-right symmetry. To achieve the left-right symmetry breaking at the scale $\Lambda_{LR}$, the Higgs fields $\omega\sim (5,\overline{5})$, $\overline{\omega}\sim (\overline{5},5)$, $\Omega \sim (10,\overline{10})$ and $\overline{\Omega}\sim (\overline{10},10)$ are introduced\footnote{Alternatively, one could break directly the left-right symmetry at the scale $\Lambda_{LR}=\Lambda$ without the need of the adjoint Higgs fields in the $(24,1)$ and $(1,24)$ representations.}. The last two steps in the pattern~\eqref{eq:pattern} are driven by the additional Higgs fields $\phi_R\sim (1,\bar{5})$, $\phi_R^c\sim (1,5)$ and $\phi_L\sim (5,1)$, $\phi_L^c \sim (\bar{5},1)$, respectively. Finally, as mentioned in the Introduction, the representations $T_{L} \sim (15,1)$, $T_{L}^c \sim (\overline{15},1)$, $T_{R} \sim (1,\overline{15})$ and $T_{R}^c \sim (1,15)$ turn out to be crucial for unification and are responsible for the Majorana masses of neutrinos.

One of the attractive features of the $\mathsf{SU(5)\times SU(5)}$ theory is the possibility of a generalized seesaw mechanism to give masses to all SM fermions through the heavy vector-like fermions~\cite{Cho:1993jb}. The Yukawa contribution to the superpotential reads as
\begin{align}
W_Y=\,\psi^{c} Y_{1}\omega\psi + \chi^{c} Y_{2}\Omega\chi +
\sqrt{2}\psi Y_{3}\chi \phi_{L}^c +
\sqrt{2}\psi^{c}Y_{3}\chi^{c}\phi_{R}^c +
\frac{1}{4}\chi Y_{4}\chi \phi_{L}+
\frac{1}{4}\chi^{c}Y_{4}\chi^{c}\phi_{R}\,,
\end{align}
where $Y_i$ denote the Yukawa coupling matrices. We choose the breaking
directions as $\left\langle \omega \right\rangle_{k}^{k}=\left\langle \Omega
\right\rangle _{12}^{12}=\left\langle \Omega
\right\rangle_{23}^{23}=\left\langle \Omega \right\rangle_{31}^{31}=\left\langle
\Omega \right\rangle_{45}^{45}=\Lambda_{LR}$, $k=1,2,3$ and
$\left\langle \phi_{L,R}\right\rangle =(0,0,0,0,v_{u\,L,R})^{T}$, $\left\langle
\phi_{L,R}^c\right\rangle =(0,0,0,0,v_{d\,L,R})^{T}$, with $v_{L,R}^2 =
v_{u\,L,R}^2+v_{d\,L,R}^2\,$. The final mass contribution to all charged
fermions can then be written as
\begin{align}\label{eq:genseesaw}
\begin{split}
-\mathcal{L}_{m}=&
\begin{pmatrix}
u & U
\end{pmatrix}
\begin{pmatrix}
0 & Y_{4}v_{u\,L} \\
Y_{4}v_{u\,R} & \,-Y_{2}\Lambda_{LR}
\end{pmatrix}
\begin{pmatrix}
u^{c} \\
U^{c}
\end{pmatrix}+
\begin{pmatrix}
d & D
\end{pmatrix}
\begin{pmatrix}
0 & Y_{3}v_{d\,L} \\
Y_{3}^{\mathsf{T}}v_{d\,R} & \,-Y_{1}\Lambda_{LR}
\end{pmatrix}
\begin{pmatrix}
d^{c} \\
D^{c}
\end{pmatrix}\,+\\
&\begin{pmatrix}
e & E
\end{pmatrix}
\begin{pmatrix}
0 & Y_{3}^{\mathsf{T}}v_{d\,L} \\
Y_{3}v_{d\,R} & \,-Y_{2}\Lambda_{LR}
\end{pmatrix}
\begin{pmatrix}
e^{c} \\
E^{c}
\end{pmatrix}.
\end{split}
\end{align}

By means of the above procedure a generalized type-I seesaw mechanism can be implemented for all light quarks and charged leptons, provided that the vector-like fermion masses, which are proportional to the $\Lambda_{LR}$ scale, are heavy enough and $v_Lv_R\ll\Lambda_{LR}^2$. As it turns out, heavy vector-like fermion masses are also required for a successful unification of gauge couplings. For the sake of simplicity, we shall assume that the breaking pattern~\eqref{eq:pattern} to the SM gauge group occurs at a unique energy scale, \emph{i.e.} $v_R\approx\Lambda_{LR}\approx\Lambda$. In this case the fermion mass spectrum has the approximate seesaw form $m_f=\mathcal{O}(y_fv_L)$ and $M_V=\mathcal{O}(y_V\Lambda_{LR})$, for light and heavy fermions, respectively. The precise realization of this generalized seesaw for fermions is beyond the scope of this work. It is our aim, instead, to discuss in detail how gauge couplings unify in this theory.

For the neutrino sector, the relevant terms in the superpotential are
\begin{equation}
W_{N}=\sqrt{2}\,Y_{5}\,(\psi \psi T_{L}+\psi^{c}\psi ^{c}T_{R}),
\end{equation}
if one assumes R-parity conservation. Then, introducing two additional supermultiplets, $(5,\overline{5})$ and $(\overline{5},5)$, with vacuum alignment in the lepton doublet direction, light neutrinos would acquire masses through the conventional (type-I and/or type-II) seesaw mechanisms. It is worth noticing that, in the absence of the Higgs multiplets $\phi_R$, $\phi_R^c$, $\phi_L$ and $\phi_L^c$, R-parity is automatically conserved\footnote{Terms in the superpotential such as $\psi\phi_L$, $\chi\phi^{c}_L\phi^{c}_L$, $T_L\psi\phi^{c}_L$ and $T_L\chi\phi_L$ violate R-parity.}~\cite{Mohapatra:1996iy}. In the latter case, quark and charged lepton masses would arise from higher dimension operators instead of the generalized seesaw Lagrangian terms given in Eq.~\eqref{eq:genseesaw}.

\section{Gauge coupling unification}

\label{sec:gut}

The two-loop renormalization group equations (RGE) for the gauge coupling constants $\alpha_i \, (i=1,2,3)$ can be written in the form
\begin{equation}
\label{eq:rge}
\frac{d}{dt}\alpha^{-1}_i=-\frac{b_i}{2\pi k_i}- \frac{1}{8\pi^2}
\sum_{j} \frac{b_{ij}\, \alpha_j}{k_i k_j}\,-\,\frac{1}{32\pi^3 k_i}\sum_{f=u,d,e}C_{if}\, \text{Tr} \left(Y_f^{\dagger}Y_f\right)\,,
\end{equation}
where $\alpha_1 = k_1\,\alpha_y,\, \alpha_2 = k_2\,\alpha_w$ and $\alpha_3 = k_3\,\alpha_s$; $b_i$ are the usual one-loop beta coefficients; $b_{ij}$ and $C_{if}$ are the two-loop beta coefficients (see Appendix~\ref{a1:beta}). The quantities $Y_f$ denote the quark and lepton Yukawa coupling matrices. At the unification scale $\Lambda$, the gauge couplings $\alpha_i$ obey the relation $\alpha_U = \alpha_1 = \alpha_2= \alpha_3$ (cf. Eq.~\eqref{unifcond}).

To get some insight into the unification in the one-loop approximation, let us define the effective beta coefficients $B_i$~\cite{Giveon:1991zm},
\begin{equation}
B_i\equiv\frac{1}{k_i}\left(b_i+\sum_I b_i^I\,r_I\right),
\end{equation}
where
\begin{equation}
r_I = \frac{\ln\left(\Lambda/M_I\right)}{\ln\left(\Lambda/M_Z\right)}\,.
\end{equation}
In the above expression, $M_I$ denotes an intermediate energy scale between the electroweak scale $M_Z$ and the GUT scale $\Lambda$, and the coefficients $b_{i}^I$ account for the new contribution to the one-loop beta functions $b_{i}$ above the threshold $M_I$. It is also convenient to introduce the differences $B_{ij}\equiv B_i-B_j$, such that
\begin{equation}
B_{ij}= B^{\text{SM}}_{ij}+\sum_I\Delta^I_{ij}r_I\,,
\end{equation}
where $B^{\text{SM}}_{ij}$ corresponds to the SM particle contribution and
\begin{equation}
\Delta^I_{ij}= \frac{b^I_i}{k_i}-\frac{b^I_j}{k_j}\,.
\end{equation}
The following $B$-test is then obtained,
\begin{equation}  \label{eq:Btest}
B\equiv\frac{B_{23}}{B_{12}}=\frac{\sin^2\theta_W-\dfrac{k_2}{k_3}\dfrac{\alpha}
{\alpha_s}}
{\dfrac{k_2}{k_1}-\left(1+\dfrac{k_2}{k_1}\right)\sin^2\theta_W}\,,
\end{equation}
together with the GUT scale relation
\begin{equation} \label{eq:Ltest}
B_{12}\, \ln \left(\frac{\Lambda}{M_Z}\right)= \frac{2\pi}{\alpha}\left[\frac{
1}{k_1}-\left(\frac{1}{k_1}+\frac{1}{k_2} \right)\sin^2\theta_W\right ].
\end{equation}

Notice that the right-hand sides of Eqs.~\eqref{eq:Btest} and~\eqref{eq:Ltest} depend only on low-energy electroweak data and the group factors $k_i$. Adopting the following experimental values at $M_Z$~\cite{Nakamura:2010zzi}
\begin{align}
\alpha^{-1}&=127.916\pm0.015\,, \\
\sin^2\theta_W&=0.23116\pm0.00013\,, \\
\alpha_s&=0.1184\pm0.0007\,,
\end{align}
the above relations read as
\begin{align}\label{eq:Btestexp}
\begin{split}
B&=0.718\pm0.003\,, \\
B_{12}\,\ln\left(\frac{\Lambda}{M_Z}\right)&=185.0\pm0.2\,,
\end{split}
\end{align}
in the canonical GUT models with $k_i=(5/3,1,1)$, \emph{e.g.} in $\mathsf{SU(5)}$ and $\mathsf{SO(10)}$. On the other hand, for the $\mathsf{SU(5)\times SU(5)}$ model where $k_i=(13/3,1,2)$ one obtains
\begin{equation}
\begin{aligned}
\label{eq:Btestexp2}
B&=-3.687\pm0.012\,, \\
B_{12}\,\ln\left(\frac{\Lambda}{M_Z}\right)&=-43.19\pm0.13\,.
\end{aligned}
\end{equation}

The coefficients $B_{ij}$ that appear in the left-hand sides of Eqs.~\eqref{eq:Btest} and~\eqref{eq:Ltest} strongly depend on the particle content of the theory. For instance, considering the SM particles with $n_H$ light Higgs doublets, one has $b_1=20/3+n_H/6$, $b_2=-10/3+n_H/6$ and $b_3=-7$, so that these coefficients are given by
\begin{equation}  \label{eq:beffSM}
B_{12}=\frac{22}{3}-\frac{n_H}{15}\,,\quad B_{23}=\frac{11}{3}+
\frac{n_H}{6}\,.
\end{equation}
In the supersymmetric case they become
\begin{equation}  \label{eq:beffMSSM}
B_{12}=\frac{22}{3}-\frac{n_H}{15}-\left(\frac43+\frac{2\,n_H}{15}\right)r_S\,,
\quad
B_{23}=\frac{11}{3}+
\frac{n_H}{6}+\left(-\frac23+\frac{n_H}{3}\right)r_S\,,
\end{equation}
with the ``running weight'' $r_S\simeq0.93$, for a low SUSY threshold $M_S \simeq 1$~TeV and a unification scale $\Lambda \simeq 10^{16}$~GeV.

It is interesting to notice that Eqs.~\eqref{eq:beffSM} and~\eqref{eq:beffMSSM} together with the constraint \eqref{eq:Btestexp} allow to determine the number of the light Higgs doublets that would be required for the unification in the canonical GUT models,
\begin{equation}
n_H=110\left(\frac{2B-1}{2B+5}\right)\approx7\quad\text{(SM)}\,,
\end{equation}
\begin{equation}
\label{eq:BtMSSM}
n_{H}=10\left(\frac{11-2r_{s}}{1+2r_{s}}\right)\left(\frac{2B-1}{2B+5}
\right) \approx2\quad\text { (MSSM)}\,.
\end{equation}
Clearly, the B-test fails badly in the SM case which possesses only one Higgs doublet, while Eq.~\eqref{eq:BtMSSM} just corroborates the fact that the gauge couplings in the MSSM seemingly unify at one-loop level. Would one take only the MSSM particle content into account, the B-test would also fail badly in the SUSY $\mathsf{SU(5)\times SU(5)}$ case. Indeed, in such a case $B\approx1.625$ which is far above the required value given in Eq.~\eqref{eq:Btestexp2} and, hence, the need for extra particles with suitable $B_{ij}$ coefficients. In Table~\ref{tab1} we present the relevant contributions $\Delta_{ij}$ to the $B_{ij}$ coefficients of the SUSY $\mathsf{SU(5)\times SU(5)}$ model which include, besides the MSSM threshold, the triplet $\Sigma_3$ and octet $\Sigma_8$ belonging to the $(24,1)$ representation, the triplets $T,T^c$ in the $(15,1)+(\overline{15},1)$ representation as well as the exotic vector-like chiral multiplets $U,U^c$, $D,D^c$ and $E,E^c$.

\begin{table}[h]
\caption{\label{tab1} The $\Delta_{ij}$ contributions to the $B_{ij}$
coefficients in the $\mathsf{SU(5)\times
SU(5)}$ case. The SM contribution to the  coefficients are
$B^{\text{SM}}_{12}=185/39$ and
$B^{\text{SM}}_{23}=1/3$.}
\centering
\begin{tabular}{cccccccc}
\\
\hline\hline
& MSSM & $\Sigma_3$ & $\Sigma_8$ & $T$ & $U$ & $D$ & $E$ \\ \hline
$\Delta_{12}$ & -125/39 & -2 & 0 & -34/13 & 24/13 & 6/13 & 18/13\\
$\Delta_{23}$ & 13/6 & 2 & -3/2 & 4 & -3/2 & -3/2 & 0 \\
\hline\hline
\end{tabular}
\end{table}

\begin{figure}[th]
\begin{centering}
\includegraphics[width=12cm]{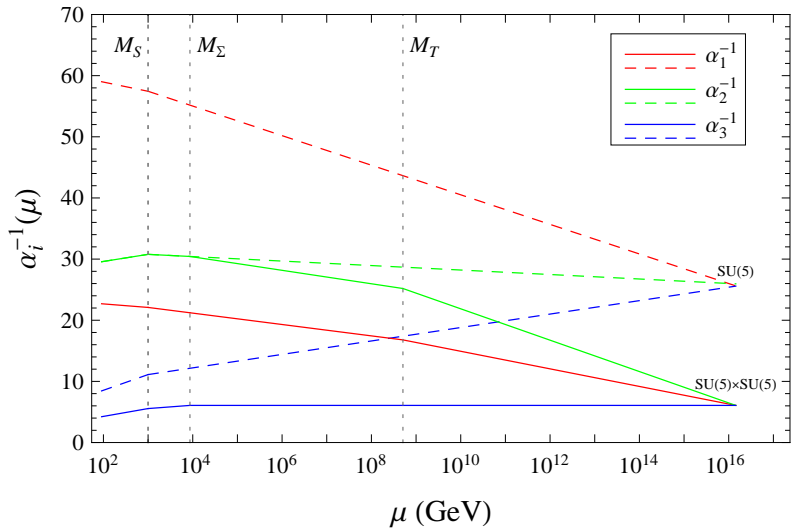}
\caption{\label{fig1} The gauge coupling running at one-loop level for the canonical $\mathsf{SU(5)}$ MSSM (dashed lines) and the $\mathsf{SU(5)\times SU(5)}$ theory (solid lines), assuming the same unification scale, $\Lambda\simeq 2\times 10^{16}$ GeV. The SUSY scale is fixed at $M_S=1$ TeV. Notice that for the non-canonical case one needs $\Sigma_3$ and $\Sigma_8$ close to $M_{\Sigma}=10$ TeV and the triplets $T,T^c$ at a higher scale near $M_T=10^9$ GeV.}
\end{centering}
\end{figure}

\begin{figure}[th]
\begin{centering}
\includegraphics[width=11cm]{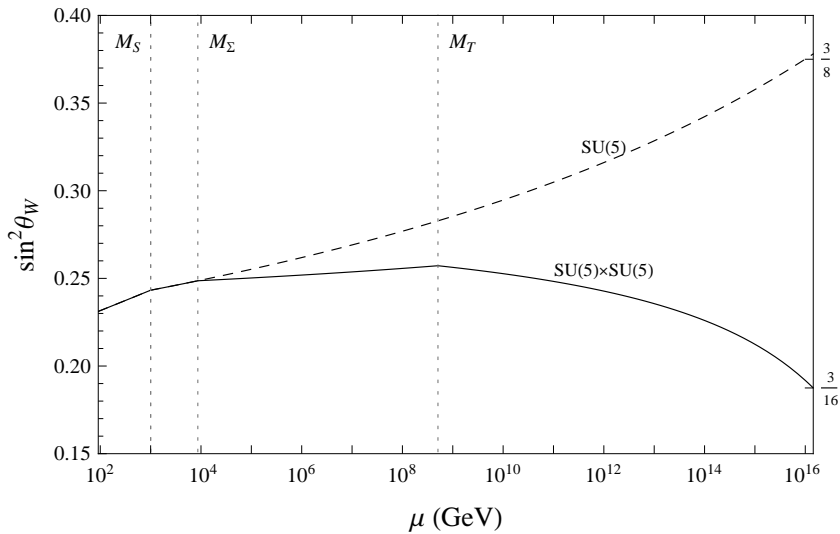}
\caption{\label{fig2} The evolution of $\mathsf{sin}^2\theta_W$ at one-loop level for the canonical $\mathsf{SU(5)}$ MSSM (dashed line) and the $\mathsf{SU(5)\times SU(5)}$ theory (solid line), assuming $\Lambda\simeq  2 \times 10^{16}$ GeV, $M_S=1$ TeV, $M_{\Sigma}=10$ TeV and $M_T=10^9$ GeV.}
\end{centering}
\end{figure}

Since Eqs.~\eqref{eq:Btestexp2} require $B_{12}<0$ and $B_{23}>0$, it becomes clear from Table~\ref{tab1} that $\Sigma_3$ and $T$ improve unification, while $U$, $D$ and $E$ act in the opposite manner and, therefore, should be heavy enough. For illustration, in Fig.~\ref{fig1} we plot the one-loop running of the gauge couplings for the SUSY $\mathsf{SU(5)}$ and $\mathsf{SU(5)\times SU(5)}$ theories, assuming a common unification scale, $\Lambda= 2 \times 10^{16}$ GeV. The SUSY threshold $M_S$ is chosen in both cases at $1$ TeV. For the $\mathsf{SU(5)\times SU(5)}$ case, we assume a common mass scale $M_{\Sigma}$ for $\Sigma_3$ and $\Sigma_8$, and for the vector-like particles $U,D,E$ we set their mass scale $M_V=\Lambda$. The one-loop unification then demands $M_{\Sigma}\simeq10$ TeV and the triplets $T,T^c$ to have a mass $M_T\simeq10^9$ GeV. The evolution of $\mathsf{sin}^2\theta_W$ at one-loop level is given in Fig.~\ref{fig2}. As antecipated in the Introduction, adding the appropriate $\mathsf{SU(5)\times SU(5)}$ representations is essential for driving the running of $\mathsf{sin}^2\theta_W$ from the low value $3/16$ at GUT scale to its correct value at the electroweak scale.

\begin{figure}[th]
\begin{centering}
\includegraphics[width=12cm]{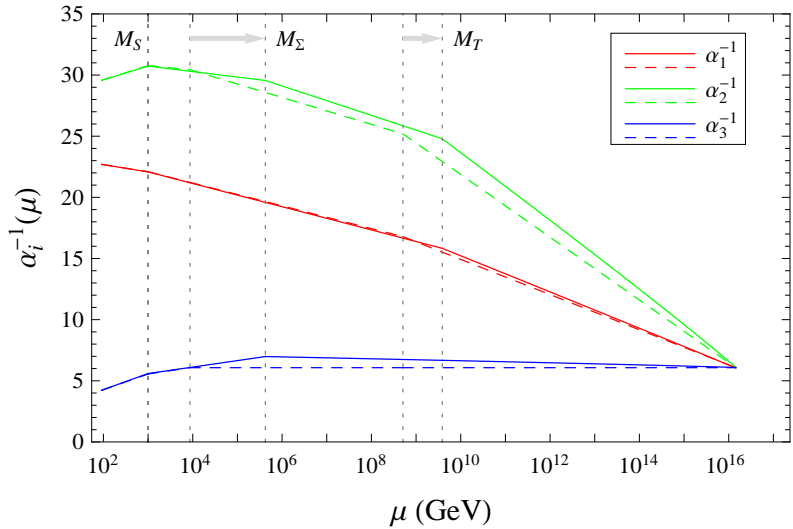}
\caption{\label{fig3} Comparison of the $\mathsf{SU(5)\times SU(5)}$ running of
gauge couplings at one-loop level (dashed lines) and two-loop level (solid lines). For a fixed
$M_S=1$ TeV and the same unification scale $\Lambda \simeq 2 \times 10^{16}$~GeV, two-loop effects increase the intermediate scales $M_\Sigma$ and $M_T$.}
\end{centering}
\end{figure}

One may wonder whether two-loop effects significantly modify the above picture. The example presented in Fig.~\ref{fig3} shows that, although the values of the gauge couplings as a function of the energy scale $\mu$ are essentially unchanged, the two-loop effects tend to increase both $M_\Sigma$ and $M_T$ scales. In the next section we shall perform a two-loop numerical analysis in order to determine the full range of the relevant intermediate mass scales.

\section{Numerical analysis}

\label{sec:num}

In this section we present a general numerical analysis of the two-loop gauge coupling unification of the $\mathsf{SU(5)\times SU(5)}$ model sketched in Sec.~\ref{sec:model}. We adopt the $\overline{\rm{DR}}$ scheme, which is appropriate for the two-loop renormalization group evolution in supersymmetric models. The measure of unification used here is given by the quantity
\begin{equation}\label{epsilon}
\epsilon =\sqrt{(\alpha _{1\Lambda}^{-1}-\alpha _{2\Lambda}^{-1})^{2}+(\alpha
_{1\Lambda}^{-1}-\alpha _{3\Lambda}^{-1})^{2}+(\alpha _{2\Lambda}^{-1}-\alpha _{3\Lambda}^{-1})^{2}}\,,
\end{equation}
which measures the ``distance'' between the couplings $\alpha
_{i\Lambda}^{-1}\equiv\alpha
_{i}^{-1}(\Lambda)$ at the
unification scale $\Lambda$.
Alternatively, one could use the quantity~\cite{Auto:2003ys}
\begin{equation}\label{Rquantity}
 R=\frac{\max(\alpha_{1\Lambda},\, \alpha_{2\Lambda},\,
\alpha_{3\Lambda})}{\min(\alpha_{1\Lambda},\,
\alpha_{2\Lambda},\,
\alpha_{3\Lambda})}\,,
\end{equation}
which measures the amount of non-unification between the largest and the smallest gauge coupling value at the scale $\Lambda$. We have verified that both quantities lead to similar unification constraints. In particular, requiring $\epsilon\lesssim 0.1$ would correspond to $R-1\lesssim0.07$.

Solving for the one-loop RGE of gauge couplings in the MSSM, and assuming a SUSY threshold $M_{S}=1$~TeV, the measure $\epsilon$ attains its minimum value, $\epsilon \simeq 0.50$, for $\Lambda \simeq 1.44 \times 10^{16}$~GeV. On the other hand, at two-loop level, its minimum is $\epsilon\simeq 0.18$ for $\Lambda\simeq 1.38 \times 10^{16}$~GeV, so that two-loop effects significantly improve unification. Inspired by the MSSM results, in our study we choose values of $\epsilon \leq 0.1$ as the criterion for unification. One can then expect that threshold effects would be sufficient to yield a perfect unification.

\begin{figure}[th]
\begin{centering}
\includegraphics[width=12cm]{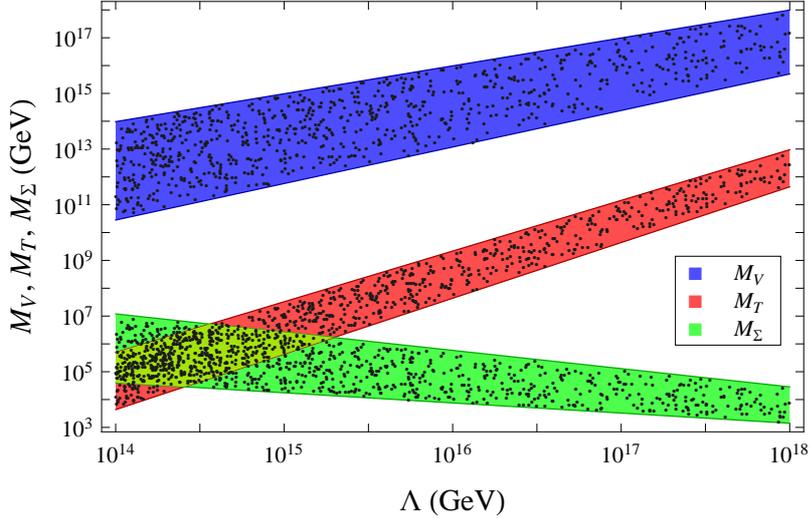}
\caption{\label{fig4}Intermediate mass scales $M_V$, $M_T$ and $M_\Sigma$ as functions of the unification scale $\Lambda$ in the $\mathsf{SU(5)\times SU(5)}$ model. The delimited color regions correspond to solutions $\alpha _{i \Lambda}^{-1}\, (i=1,2,3)$ with a unification measure $\epsilon \leq 0.1$ at two loops.}
\end{centering}
\end{figure}

We proceed to integrate numerically the two-loop RGEs in Eqs.~\eqref{eq:rge} from the electroweak scale $M_Z$ to a randomly chosen unification scale $\Lambda\gtrsim10^{14}$ GeV. The intermediate vector-like fermion mass scale $M_V$, and that of the triplet scalar, $M_T$, as well as the common scale $M_{\Sigma}$ for $\Sigma_3$ and $\Sigma_8$, are also randomly taken. The SUSY threshold scale is fixed at $M_{S}=1$~TeV. At two-loop level, the parameter space for the three relevant quantities, $M_{V}$, $M_{T}$, and $M_{\Sigma}$, is given as a function of the unification scale $\Lambda $ in Fig.~\ref{fig4}. We notice that every point corresponds to a different solution which has passed the criterion $\epsilon \leq 0.1$. As can be easily seen from the figure, the triplet mass scale $M_T$ can be close to the SUSY breaking mass scale $M_S$ for a low unification scale $\Lambda \simeq 10^{14}$~GeV. As $\Lambda$ increases, the value of $M_T$ also increases. We find $4.5\times10^3\,\text{GeV}\lesssim M_T\lesssim\,1.2\times10^{13}\,\text{GeV}$ for $10^{14}\,\text{GeV}\lesssim \Lambda\lesssim10^{18}\,\text{GeV}$. In contrast, the common mass scale $M_{\Sigma}$ decreases smoothly as $\Lambda$ increases and, for $\Lambda \sim 10^{18}$~GeV, can be as low as 1~TeV. The allowed mass range is $1.2\times10^3\,\text{GeV} \lesssim M_{\Sigma}\lesssim\,2.7\times10^{7}\,\text{GeV}$. We also note that, when $\Lambda \simeq 10^{14-15}$~GeV, both mass scales, $M_T$ and $M_{\Sigma}$, can be of the same order of magnitude. When compared to other intermediate states, vector-like fermions require a much higher mass scale. For $\Lambda \simeq 10^{14}$~GeV, we find the lower bound $M_V\gtrsim 3.2 \times 10^{10}$ GeV, while for $\Lambda \simeq 10^{17}$~GeV this bound gets more restrictive, $M_V\gtrsim10^{15}$~GeV.

\begin{figure}[th]
\begin{centering}
\includegraphics[width=12cm]{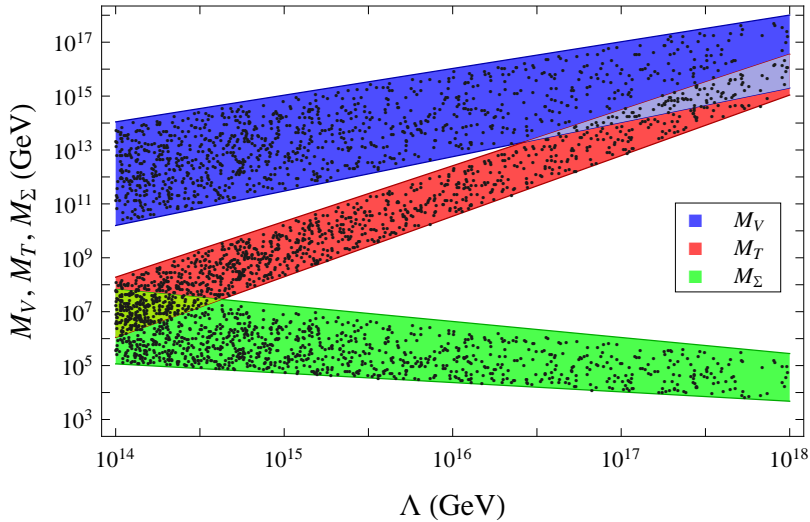}
\caption{\label{fig5} As in Fig.~\ref{fig4}, but for the $\mathsf{SU(5)\times SU(5)}$ model with two additional Higgs doublets at an intermediate scale. Notice that the triplet mass scale $M_T$ is significantly larger, and reaches values of the order of the exotic fermion mass scale $M_V$ at a high unification scale.}
\end{centering}
\end{figure}

We have also verified how sensitive the results are with respect to the variation of the SUSY breaking mass scale. In fact, no significant changes occur and the variation of the SUSY mass scale in the interval $M_S=1-100$ TeV leads only to a slight dispersion of $M_{\Sigma}$ towards lower values. No relevant modification is either observed for the parameters in Fig.~\ref{fig4}, if one considers a splitting between the masses of the triplet $\Sigma_3$ and octet $\Sigma_8$. Motivated by the rich scalar structure, we have also looked for solutions when two additional Higgs doublets are randomly inserted at some new threshold, $M_{H}$. The effects of the latter on the mass scales $M_{V}$, $M_{T}$ and $M_{\Sigma }$, given as a function of the unification scale $\Lambda$, are shown in Fig.~\ref{fig5}. While the inclusion of the two additional Higgs doublets does not significantly affect the parameter region of $M_V$ and $M_{\Sigma}$, it is clear from Fig.~\ref{fig5} that the triplet mass scale $M_T$ is shifted to much higher values, bringing $M_T$ close to the vector-like fermion mass scale for $\Lambda\gtrsim10^{16}$~GeV. The allowed ranges for the relevant scales are now given by $1.4\times10^{10}\,\text{GeV} \lesssim M_{V}\lesssim\,9.7\times10^{17}\,\text{GeV}$, $9.5\times 10^5\,\text{GeV} \lesssim M_{T}\lesssim\,4.4\times10^{16}\,\text{GeV}$, and $3.9\times10^3\,\text{GeV} \lesssim M_{\Sigma}\lesssim\,1.4\times10^{8}\,\text{GeV}$, with $M_H$ varying in the interval $10^3 \,\text{GeV} \lesssim M_{H}\lesssim\,9.4\times10^{17}\,\text{GeV}$.

\begin{figure}[h!tp]
\begin{centering}
\includegraphics[width=10cm]{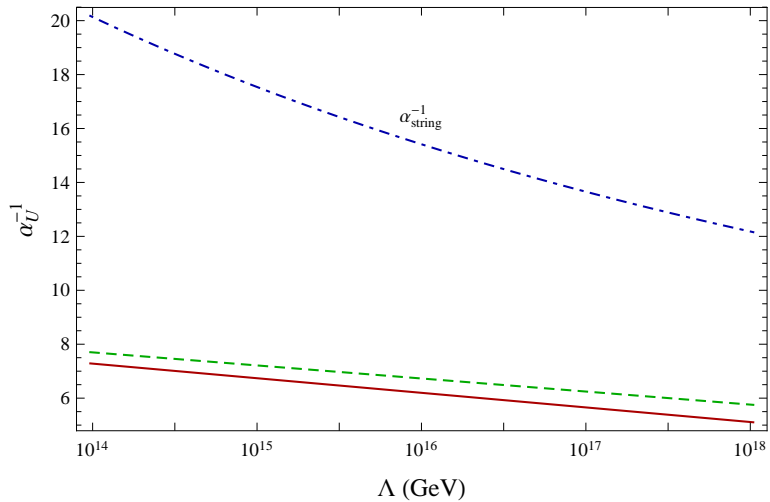}
\caption{\label{fig6} Upper values of $\alpha_{U}^{-1}$ at two-loop level in the
$\mathsf{SU(5)\times SU(5)}$ model. The solid line corresponds to the
$\mathsf{SU(5)\times SU(5)}$ model, assuming only two light Higgs doublets,
while the dashed line corresponds to the case where two extra Higgs doublets
are introduced at some intermediate scale.}
\end{centering}
\end{figure}

From the above results it becomes clear that the unification scale $\Lambda$ can reach and even exceed the perturbative string scale, $\Lambda_{s} \simeq 5.27\times 10^{17}$~GeV~\cite{Kaplunovsky:1987rp,Kaplunovsky:1992vs}. It is well known that $\mathsf{SU(5)\times SU(5)}$ theories can be embedded in the heterotic string context~\cite{Gross:1984dd,Gross:1985fr,Gross:1985rr, Barbieri:1994jq,Maslikov:1996gn,Kakushadze:1997ne}. Furthermore, in a minimal string-scale unification setup with vector-like fermions, it is conceivable to have unification of gauge couplings and gravity at the weakly coupled heterotic string scale~\cite{EmmanuelCosta:2005nh}. We may ask ourselves whether it is possible to achieve such a unification in the $\mathsf{SU(5)\times SU(5)}$ framework under consideration. In the heterotic string scenario, an additional constraint on the gauge couplings must be verified at the string scale $\Lambda_s$,
\begin{align}
\label{eq:astring}
\alpha_U=\alpha_\text{string}=\frac1{4\pi}
\left(\frac{\Lambda}{\Lambda_{s}}\right)^{2}.
\end{align}
Requiring $\Lambda\leq\Lambda_s$ in order to be in the perturbative regime, the constraint in Eq.~\eqref{eq:astring} clearly implies a lower bound on the unified gauge coupling, namely, $\alpha_{U}^{-1}\geq 4\pi$. In Fig.~\ref{fig6}, we present the upper values of $\alpha_{U}^{-1} \simeq \alpha_{2 \Lambda}^{-1}$ as a function of the unification scale $\Lambda$, together with the corresponding values of $\alpha_\text{string}^{-1}$. We conclude that string unification cannot be achieved, since $\alpha_{U}^{-1}$ is very small compared to the required value of $\alpha_\text{string}^{-1}$. This conclusion also remains valid when two additional Higgs doublets are included at an intermediate energy scale.

\section{Conclusions}
\label{sec:Conclusions}

We have investigated the possibility to achieve unification of the SM gauge couplings in the context of a SUSY $\mathsf{SU(5)}_{L}\times \mathsf{SU(5)}_{R}$ GUT. For a successful gauge coupling unification, the inclusion of $(\overline{15},1)+(1,15)$ and their conjugates $(15,1)+(1,\overline{15})$ at an intermediate scale $M_T$ was essential to drive $\sin^2\theta_W$ to the correct value at the electroweak scale. From the two-loop numerical analysis, we have found that the intermediate mass scales $M_T$, $M_{\Sigma}$ for $\Sigma_3$, $\Sigma_8$ and $M_V$ for the vector-like fermions must be properly chosen to guarantee unification at the required level. As it can be clearly seen from Figs.~\ref{fig4} and~\ref{fig5}, there is a wide region allowed for these mass scales.

Models based on $\mathsf{SU(5)}_{L}\times \mathsf{SU(5)}_{R}$ unification
enclose many attractive features. Compared with the standard $\mathsf{SU(5)}$
GUT, proton decay via dimension-six operators through heavy lepto-quark
gauge bosons is suppressed, since at tree level the latter do not mediate
transitions involving only light fermions. On the other hand, the presence of
the color Higgs triplets $H_C^L$ and $H_C^R$, contained in the chiral
super-quintets $\phi_L$, $\phi_L^c$, $\phi_R$ and $\phi_R^c$, may induce proton
decay through dimension-five operators. Indeed, proton decay arises in the
lowest order from the operators $\chi\chi\chi\psi$ and
$\chi^c\chi^c\chi^c\psi^c$, which lead to the effective operators $QQQL$ with
coefficients proportional to $Y_3Y_4/M_{H_C}$, for both left and right light
matter fields. This requires that the mass scales of left and right color Higgs
triplets should be heavy enough, thus constraining the unification
scale~\cite{Nath:2006ut}. In the absence of the fields $\phi_{L,R}$ and
$\phi_{L,R}^c$ not only proton is stable at the renormalizable level, but also
R-parity is automatically conserved~\cite{Mohapatra:1996iy}. R-parity invariance
is an appealing feature in SUSY theories, since the lightest supersymmetric
particle is absolutely stable, thus providing a natural cold dark matter
candidate.

Finally, we have shown that, in the minimal $\mathsf{SU(5)}_{L}\times \mathsf{SU(5)}_{R}$ setup considered, it is not possible to achieve the unification of the gauge couplings with the gravitational coupling at the perturbative heterotic string scale. It would be interesting to investigate whether the inclusion of additional representations could help in bringing into agreement the four couplings.

\section*{Acknowledgements}

This work was partially supported by Funda\c{c}\~{a}o para a Ci\^{e}ncia e a Tecnologia (FCT, Portugal) through the Grant No. CERN/FP/109305/2009 and the project CFTP-FCT UNIT 777, which are partially funded through POCTI (FEDER). The work of D.E.C. was also supported by Accion Complementaria Luso-Espanhola FCT and MICINN with project number 20NSML3700. E.T.F. was partially supported by CNPq through the Grant No. 150416/2011-3 and by the EU project MRTN-CT-2006-035505.

\appendix

\section{One-loop and two-loop beta coefficients}

\label{a1:beta}
In this appendix we collect the $\beta$-function coefficients for the relevant particle content of the $\mathsf{SU(5)}\times\mathsf{SU(5)}$ theory.

Below the SUSY threshold $M_S$, the $\beta$-function coefficients are those
of the SM:
\begin{align}
b_i=\begin{pmatrix}
41/6 & -19/6 & -7\end{pmatrix}\,,\quad
b_{ij}=\begin{pmatrix}
{199}/{18} & 9/2 & {44}/{3} \\
3/2 & 35/6 & 12 \\
{11}/{6} & 9/2 & -26
\end{pmatrix}.
\end{align}
Above $M_S$, the coefficients are the usual MSSM ones:
\begin{align}
b_i=\begin{pmatrix}
11 & 1 & -3\end{pmatrix}, \quad
b_{ij}=\begin{pmatrix}
{199}/{9} & 9 & {88}/{3} \\
3 & 25 & 24 \\
{11}/{3} & 9 & 14
\end{pmatrix}.
\end{align}
The two-loop coefficients $C_{if}$ that account for the Yukawa contributions are
\begin{align}
C_{if}=
\begin{pmatrix}
{26}/{3} & {14}/{3} & 6 \\
6 & 6 & 2 \\
4 & 4 & 0
\end{pmatrix}.
\end{align}
We have also the following coefficients for the triplet $\Sigma_3$,
the octet $\Sigma_8$, the triplet $T$, and the vector-like fermions $U$, $D$
and $E$:
\begin{align}
b_i^{\Sigma_3}&=\begin{pmatrix}
0 & 2 & 0\end{pmatrix}\,,\quad
b_{ij}^{\Sigma_3}=\begin{pmatrix}
0 & 0 & 0\\
0 & 24 & 0\\
0 & 0 & 0
\end{pmatrix},
\\
b_i^{\Sigma_8}&=\begin{pmatrix}
0 & 0 & 3\end{pmatrix}\,,
\quad
b_{ij}^{\Sigma_8}=\begin{pmatrix}
0 & 0 & 0\\
0 & 0 & 0\\
0 & 0 & 54
\end{pmatrix},
\\
b_i^{T}&=\begin{pmatrix}
6 & 4 & 0\end{pmatrix}
\,,\quad
b_{ij}^{T}=\begin{pmatrix}
24 & 48 & 0\\
16 & 48 & 0\\
0 & 0 & 0
\end{pmatrix},
\\
b_i^{U}&=\begin{pmatrix}
8 & 0 & 3\end{pmatrix}
\,,\quad
b_{ij}^{U}=\begin{pmatrix}
128/9 & 0 & 128/3\\
0 & 0 & 0\\
16/3 & 0 & 34
\end{pmatrix},
\\
b_i^{D}&=\begin{pmatrix}
2 & 0 & 3\end{pmatrix}
\,,\quad
b_{ij}^{D}=\begin{pmatrix}
8/9 & 0 & 32/3\\
0 & 0 & 0\\
4/3 & 0 & 34
\end{pmatrix},
\\
b_i^{E}&=\begin{pmatrix}
6 & 0 & 0\end{pmatrix}
\,,\quad
b_{ij}^{E}=\begin{pmatrix}
24 & 0 & 0\\
0 & 0 & 0\\
0 & 0 & 0
\end{pmatrix},
\end{align}
which are introduced at the appropriate intermediate scales.


\begin{thebibliography}{99}

\bibitem{Georgi:1974sy}
H.~Georgi and S.~Glashow, {\it {Unity of All Elementary Particle Forces}},
  {\em Phys. Rev. Lett.} {\bf 32} (1974) 438--441.

\bibitem{Georgi:1975qb}
H.~Georgi, {\it {Unified Gauge Theories}},  {\em In Coral Gables 1975,
  Proceedings, Theories and Experiments In High Energy Physics, New York}
  (1975) 329--339.

\bibitem{Langacker:1980js}
P.~Langacker, {\it {Grand Unified Theories and Proton Decay}},  {\em Phys.
  Rept.} {\bf 72} (1981) 185.

\bibitem{Raby:2008gh}
S.~Raby, {\it {SUSY GUT Model Building}},  {\em Eur. Phys. J. C} {\bf 59}
  (2009) 223--247, [\href{http://xxx.lanl.gov/abs/0807.4921}{{\tt 0807.4921}}].

\bibitem{Sakai:1981gr}
N.~Sakai, {\it {Naturalness in Supersymmetric Guts}},  {\em Z. Phys. C} {\bf
  11} (1981) 153.

\bibitem{Barger:1992ac}
V.~D. Barger, M.~Berger, and P.~Ohmann, {\it {Supersymmetric grand unified
  theories: Two loop evolution of gauge and Yukawa couplings}},  {\em Phys.
  Rev. D} {\bf 47} (1993) 1093--1113,
  [\href{http://xxx.lanl.gov/abs/hep-ph/9209232}{{\tt hep-ph/9209232}}].

\bibitem{Martens:2010nm}
W.~Martens, L.~Mihaila, J.~Salomon, and M.~Steinhauser, {\it {Minimal
  Supersymmetric SU(5) and Gauge Coupling Unification at Three Loops}},  {\em
  Phys. Rev. D} {\bf 82} (2010) 095013,
  [\href{http://xxx.lanl.gov/abs/1008.3070}{{\tt arXiv:1008.3070}}].

\bibitem{Buras:1977yy}
A.~Buras, J.~R. Ellis, M.~Gaillard, and D.~V. Nanopoulos, {\it {Aspects of the
  Grand Unification of Strong, Weak and Electromagnetic Interactions}},  {\em
  Nucl. Phys. B} {\bf 135} (1978) 66--92.

\bibitem{Sakai:1981pk}
N.~Sakai and T.~Yanagida, {\it {Proton Decay in a Class of Supersymmetric Grand
  Unified Models}},  {\em Nucl. Phys. B} {\bf 197} (1982) 533.

\bibitem{Hayato:1999az}
Super-Kamiokande Collaboration, Y.~Hayato {\em et.~al.}, {\it {Search for
  proton decay through p $\rightarrow$ anti-neutrino K+ in a large water
  Cherenkov detector}},  {\em Phys. Rev. Lett.} {\bf 83} (1999) 1529--1533,
  [\href{http://xxx.lanl.gov/abs/hep-ex/9904020}{{\tt hep-ex/9904020}}].

\bibitem{Nath:2006ut}
P.~Nath and P.~Fileviez~Perez, {\it {Proton stability in grand unified
  theories, in strings and in branes}},  {\em Phys. Rept.} {\bf 441} (2007)
  191--317, [\href{http://xxx.lanl.gov/abs/hep-ph/0601023}{{\tt
  hep-ph/0601023}}].

\bibitem{Pati:1973rp}
J.~C. Pati and A.~Salam, {\it {Is Baryon Number Conserved?}},  {\em Phys. Rev.
  Lett.} {\bf 31} (1973) 661--664.

\bibitem{Pati:1974yy}
J.~C. Pati and A.~Salam, {\it {Lepton Number as the Fourth Color}},  {\em Phys.
  Rev. D} {\bf 10} (1974) 275--289.

\bibitem{Davidson:1987mi}
A.~Davidson and K.~C. Wali, {\it {$SU(5)_L \times SU(5)_R$ hybrid
  unification}},  {\em Phys. Rev. Lett.} {\bf 58} (1987) 2623.

\bibitem{Dine:2002se}
M.~Dine, Y.~Nir, and Y.~Shadmi, {\it {Product groups, discrete symmetries, and
  grand unification}},  {\em Phys. Rev. D} {\bf 66} (2002) 115001,
  [\href{http://xxx.lanl.gov/abs/hep-ph/0206268}{{\tt hep-ph/0206268}}].

\bibitem{Barr:1996kp}
S.~M. Barr, {\it {The Stability of the gauge hierarchy in SU(5) $\times$
  SU(5)}},  {\em Phys. Rev. D} {\bf 55} (1997) 6775--6779,
  [\href{http://xxx.lanl.gov/abs/hep-ph/9607359}{{\tt hep-ph/9607359}}].

\bibitem{Mohapatra:1996fu}
R.~Mohapatra, {\it {SU(5) $\times$ SU(5) unification model with see-saw
  mechanism and automatic R-parity conservation}},  {\em Phys. Lett. B} {\bf
  379} (1996) 115--120, [\href{http://xxx.lanl.gov/abs/hep-ph/9601203}{{\tt
  hep-ph/9601203}}].

\bibitem{Mohapatra:1996iy}
R.~Mohapatra, {\it {SU(5)$\times$SU(5) unification and automatic R-parity
  conservation}},  {\em Phys. Rev. D} {\bf 54} (1996) 5728--5733.

\bibitem{Cho:1993jb}
P.~L. Cho, {\it {Unified universal seesaw models}},  {\em Phys. Rev. D} {\bf
  48} (1993) 5331--5341, [\href{http://xxx.lanl.gov/abs/hep-ph/9304223}{{\tt
  hep-ph/9304223}}].

\bibitem{Giveon:1991zm}
A.~Giveon, L.~J. Hall, and U.~Sarid, {\it {SU(5) unification revisited}},  {\em
  Phys. Lett. B} {\bf 271} (1991) 138--144.

\bibitem{Nakamura:2010zzi}
Particle Data Group Collaboration, K.~Nakamura {\em et.~al.}, {\it
  {Review of particle physics}},  {\em J. Phys. G} {\bf 37} (2010) 075021.

\bibitem{Auto:2003ys}
D.~Auto {\em et.~al.}, {\it {Yukawa coupling unification in supersymmetric
  models}},  {\em JHEP} {\bf 06} (2003) 023,
  [\href{http://xxx.lanl.gov/abs/hep-ph/0302155}{{\tt hep-ph/0302155}}].

\bibitem{Kaplunovsky:1987rp}
V.~S. Kaplunovsky, {\it {One Loop Threshold Effects in String Unification}},
  {\em Nucl. Phys. B} {\bf 307} (1988) 145,
  [\href{http://xxx.lanl.gov/abs/hep-th/9205068}{{\tt hep-th/9205068}}].

\bibitem{Kaplunovsky:1992vs}
V.~S. Kaplunovsky, {\it {One loop threshold effects in string unification}},
  \href{http://xxx.lanl.gov/abs/hep-th/9205070}{{\tt hep-th/9205070}}.

\bibitem{Gross:1984dd}
D.~J. Gross, J.~A. Harvey, E.~J. Martinec, and R.~Rohm, {\it {The Heterotic
  String}},  {\em Phys. Rev. Lett.} {\bf 54} (1985) 502--505.

\bibitem{Gross:1985fr}
D.~J. Gross, J.~A. Harvey, E.~J. Martinec, and R.~Rohm, {\it {Heterotic String
  Theory. 1. The Free Heterotic String}},  {\em Nucl. Phys. B} {\bf 256} (1985)
  253.

\bibitem{Gross:1985rr}
D.~J. Gross, J.~A. Harvey, E.~J. Martinec, and R.~Rohm, {\it {Heterotic String
  Theory. 2. The Interacting Heterotic String}},  {\em Nucl. Phys. B} {\bf 267}
  (1986) 75.

\bibitem{Barbieri:1994jq}
R.~Barbieri, G.~R. Dvali, and A.~Strumia, {\it {Strings versus supersymmetric
  GUTs: Can they be reconciled?}},  {\em Phys. Lett. B} {\bf 333} (1994)
  79--82, [\href{http://xxx.lanl.gov/abs/hep-ph/9404278}{{\tt
  hep-ph/9404278}}].

\bibitem{Maslikov:1996gn}
A.~Maslikov, I.~Naumov, and G.~Volkov, {\it {The Paths of unification in the
  GUST with the G x G gauge groups of E(8) $\times$ E(8)}},  {\em Phys. Lett.
  B} {\bf 409} (1997) 160--172,
  [\href{http://xxx.lanl.gov/abs/hep-th/9612243}{{\tt hep-th/9612243}}].

\bibitem{Kakushadze:1997ne}
Z.~Kakushadze and S.~Tye, {\it {A Classification of three family grand
  unification in string theory. 2. The SU(5) and SU(6) models}},  {\em Phys.
  Rev. D} {\bf 55} (1997) 7896--7908,
  [\href{http://xxx.lanl.gov/abs/hep-th/9701057}{{\tt hep-th/9701057}}].

\bibitem{EmmanuelCosta:2005nh}
D.~Emmanuel-Costa and R.~Gonz\'{a}lez~Felipe, {\it {Minimal string-scale
  unification of gauge couplings}},  {\em Phys. Lett. B} {\bf 623} (2005)
  111--118, [\href{http://xxx.lanl.gov/abs/hep-ph/0505257}{{\tt
  hep-ph/0505257}}].

\end{thebibliography}
\end{document}